\documentclass[prd,twocolumn,nofootinbib]{revtex4}
\usepackage{graphicx, epsfig}
\usepackage{color}
\usepackage{mathrsfs}
\usepackage{bm}
\usepackage{amsmath,amssymb}
\usepackage[caption=false]{subfig}
\usepackage{hyperref}
\usepackage[normalem]{ulem}
\usepackage{mathtools}

\newcommand{\be}{\begin{equation}}
\newcommand{\ee}{\end{equation}}
\newcommand{\ba}{\begin{eqnarray}}
\newcommand{\ea}{\end{eqnarray}}

\begin{document}
\title{A Classical-Quantum Correspondence and Backreaction}
\author{Tanmay Vachaspati$^*$, George Zahariade$^{*\dag}$}
\affiliation{
$^*$Physics Department, Arizona State University, Tempe, AZ 85287, USA. \\
$^\dag$Beyond Center for Fundamental Concepts in Science, Arizona State University, Tempe, AZ 85287, USA. \\
}

\begin{abstract}
\noindent
We work in the Heisenberg picture to demonstrate the classical-quantum correspondence (CQC) in which 
the dynamics of a quantum variable is equivalent to that of a complexified classical variable.
The correspondence provides a tool for analyzing quantum backreaction problems which we illustrate by a toy 
model in which a rolling particle slows down due to quantum radiation. The dynamics found using the CQC is 
in excellent agreement with that found using the much more laborious full quantum analysis. 
\end{abstract}

\maketitle

A large class of physical systems involve classical dynamics that is coupled to quantum 
degrees of freedom that get excited as the classical system evolves. Examples of such 
systems include particle production during cosmological evolution,  Hawking radiation 
during gravitational collapse and Schwinger pair production in an electric field. The key 
question we address in this paper is: how do we account for the backreaction of the
quantum excitations on the classical background? The question is of fundamental 
interest as its solution may hold the key to many problems of current interest including 
the black hole information paradox.

%

Past work on the backreaction question is usually framed as a
perturbative-iterative process; the radiation is calculated in perturbation theory,
the backreaction is then calculated semiclassically, which then leads to modified
radiation, and so on. In the present work, we instead develop a classical-quantum
correspondence (CQC) using which we can transform the quantum radiation problem
into a classical radiation problem. Then the entire problem, including backreaction, 
can be cast as a set of classical equations with definite initial 
conditions~\cite{Vachaspati:2017jtw,Vachaspati:2018pps}. 
These equations can then be solved numerically. 
(Other work on classical-quantum connections
includes \cite{Brout:1993be,Halliwell:1997hy,Anderson:1994si,Cooper:1989kf,Hertzberg:2016tal}.)

The system we have in mind consists of a classical background variable
that couples to a free quantum field. Expanding the field in modes, the mode coefficients 
behave like an infinite set of simple harmonic oscillators with time-dependent mass and 
frequency. 
By redefining variables, it is possible to eliminate the time-dependence of the mass.
Thus the field
theory problem can be mapped into a quantum mechanics problem consisting
of an infinite set of simple harmonic oscillators with time-dependent frequencies 
determined by the dynamics of the background variable.

The quantum simple harmonic oscillator (qSHO) with a time-dependent frequency
has been solved in terms of a two-dimensional classical SHO (cSHO) in early 
work~\cite{Lewis:1968yx,Lewis:1968tm,doi:10.1119/1.1986050} and more
recently~\cite{Jones-Smith:2018lxs}.
In Ref.~\cite{Vachaspati:2018pps}, we have used this connection in the Schr\"odinger 
picture to show the equivalence of the quantum and classical systems,
and we have discussed the excitations produced due to the time-dependence of
the frequency.
This result can 
be applied to a mode by mode analysis of a free quantum field to show that the 
quantum field dynamics 
can be described in terms of the classical dynamics of a corresponding system
with prescribed initial conditions \cite{Vachaspati:2018hcu}.
However, particle production is usually discussed in the Heisenberg picture
using the method of Bogoliubov transformations ({\it e.g.} \cite{Birrell:1982ix})
while functional Schr\"odinger derivations are less familiar. 
In this paper we first close this gap by demonstrating the classical-quantum
correspondence (CQC) in the Heisenberg picture. 

The production of particles due to the time varying background will cause 
dissipation in the time variation. The CQC provides a simple tool to study this
backreaction because quantum dynamics can be replaced by classical dynamics.
To assess the validity of the CQC approach to backreaction, we construct a
toy model in which we can find the backreaction using the CQC and also in
the full quantum theory. The results are in excellent agreement and the accuracy
of the CQC approach increases as the background becomes more classical.

This paper is organized as follows.
We first show the CQC between a qSHO and two cSHOs in the Heisenberg
picture when the frequency of the SHOs is an arbitrarily varying function of time. 
This is done in three steps. First, in Sec.~\ref{heisenberg} we derive the Heisenberg
equations of motion for the ladder operators with a time-dependent frequency.
Then in Sec.~\ref{bogo} we find the energy radiated in particles by the method of Bogoliubov 
transformations. In Sec.~\ref{classicalbogo} we show that the dynamics of the
radiation and, in particular, the energy in quantum radiation, can be found by a purely 
classical calculation that involves doubling the radiative degrees of freedom, or
equivalently complexifying these degrees of freedom. Having thus established
the CQC, we turn to the quantum radiation backreaction on the 
dynamics of the classical variable. We find the results obtained using the
CQC and compare them to the full quantum dynamics that are found by
using novel, though laborious, numerical methods described in the appendix. 
Our conclusions are given in Sec.~\ref{conclusions}.

\section{CQC in Heisenberg Picture}
\label{cqc}

\subsection{Heisenberg equations}
\label{heisenberg}

The Hamiltionian for a simple harmonic oscillator with time-dependent frequency is
\be
H = \frac{p^2}{2m} + \frac{m\omega^2}{2} x^2
\label{Ham}
\ee
where $\omega = \omega(t)$ is an unspecified function. We define ladder operators in
the usual way
\be
a = \frac{p-im\omega x}{\sqrt{2m\omega}}, \ \ 
a^\dag = \frac{p+im\omega x}{\sqrt{2m\omega}}
\label{aadagdefn}
\ee
It is straight-forward to check that $[a,a^\dag]=1$ even for a time-dependent $\omega$.
Then,
\be
H = \omega (t) \left ( a^\dag a + \frac{1}{2} \right )
\ee
and
\be
\frac{\partial a}{\partial t} = - \frac{\dot \omega}{2\omega} a^\dag, \ \ 
\frac{\partial a^\dag}{\partial t} = - \frac{\dot \omega}{2\omega} a
\ee

We now go to the Heisenberg picture. Then the equation of motion for $a$ is
\ba
\frac{da}{dt} &=& -i [a,H] + \frac{\partial a}{\partial t} \nonumber \\
&=& -i\omega a - \frac{\dot \omega}{2\omega} a^\dag
\label{aeq}
\ea
and similarly
\be
\frac{da^\dag}{dt} = +i\omega a^\dag - \frac{\dot \omega}{2\omega} a
\label{adageq}
\ee

\subsection{Bogoliubov transformation}
\label{bogo}

To obtain the excitation of the simple harmonic oscillator due to the time-dependence
of $\omega$, we write
\be
a(t) = \alpha(t) a_0 + \beta (t) a_0^\dag , \ \ 
a^\dag (t) = \alpha^*(t) a^\dag_0 + \beta^* (t) a_0
\ee
where $a_0$ and $a_0^\dag$ are the annihilation and creation operators in
Eq.~(\ref{aadagdefn}) at the initial time, $t=0$.
%
The commutation relation $[a,a^\dag]=1$ leads to the constraint
\be
|\alpha|^2 - | \beta|^2 =1.
\label{conserved}
\ee
and Eqs.~(\ref{aeq}), (\ref{adageq}) lead to
\ba
{\dot \alpha} &=& - i \omega \alpha - \frac{\dot \omega}{2\omega} \beta^* 
\label{alphaeq} \\
{\dot \beta} &=& - i \omega \beta - \frac{\dot \omega}{2\omega} \alpha^*
\label{betaeq}
\ea
These equations also lead to the constraint
\be
\beta {\dot \alpha} - \alpha {\dot \beta} = \frac{\dot \omega}{2\omega}.
\ee
The expectation value of the energy in the vacuum state is
\be
E_q(t) \equiv \langle H \rangle = \omega(t) \left ( |\beta |^2 + \frac{1}{2} \right ).
\label{Eq}
\ee

\subsection{The CQC}
\label{classicalbogo}

Here we show that the quantum dynamics of the time-dependent simple
harmonic oscillator is given by the classical dynamics of two classical
simple harmonic oscillators if we impose certain initial conditions. In contrast 
to the earlier derivation in the functional Schr\"odinger 
picture~\cite{Vachaspati:2006ki,Kolopanis:2013sty,Vachaspati:2018pps}, here 
we show this 
correpondence in the Heisenberg picture. Particle production is then given by 
the Bogoliubov transformation method, and the energy in quantum excitations, 
$E_q$, is also identical to the energy of two classical simple harmonic oscillators.


We rewrite Eqs.~(\ref{alphaeq}) and (\ref{betaeq}) as a single second order equation by doing the change of variables
\ba
\alpha &=& \sqrt{\frac{m}{2\omega}} \left ( {\dot z}^* - i \omega z^* \right )\,,
\label{alphasoln} \\
\beta &=& \sqrt{\frac{m}{2\omega}} \left ( {\dot z} -  i \omega z \right )\,,
\label{betasoln}
\ea
where
\be
z \equiv \xi + i \chi
\ee
is complex and $\xi$ and $\chi$ are its real and imaginary parts. The overall factor of $\sqrt{m}$
in Eqs.~(\ref{alphasoln}) and (\ref{betasoln}) ensures that $\alpha$ and $\beta$ have the correct mass 
dimensions equal to zero when $z$ has dimensions of length.

The expressions for $\alpha$ and $\beta$ are identical to the definition
of the annihilation operator $a$ in Eq.~(\ref{aadagdefn}) if we think of $z$
and $m{\dot z}$ as representing the complexified position and momentum 
operators for one dynamical variable and similarly $z^*$ and $m{\dot z}^*$
for a second dynamical variable. 
Similarly the complex conjugates $\alpha^*$ and $\beta^*$ then correspond 
to the expression for the creation operator $a^\dag$ in Eq.~(\ref{aadagdefn}).

Inserting Eqs.~(\ref{alphasoln}) and (\ref{betasoln}) in Eqs.~(\ref{alphaeq}) 
and (\ref{betaeq}) we find the equation of motion satisfied by $z$,
\be
{\ddot z} + \omega^2(t) z =0\,,
\label{zeq}
\ee
which describes the dynamics of a two-dimensional cSHO ({\it e.g.} a point mass attached to a rotating spring on a plane) with time dependent frequency, or equivalently, two such one-dimensional cSHOs.
Hence $\xi$ and $\chi$ (the two-dimensional Cartesian coordinates of the point mass) satisfy the classical equations of motion
\be
{\ddot \xi}+\omega^2(t) \xi =0, \ \ {\ddot \chi}+\omega^2(t) \chi =0.
\label{xichieq}
\ee

The initial condition, $a(0)=a_0$, corresponds to: $\alpha(0)=1$, $\beta (0)=0$, 
which imply
\be
z(0) = \frac{-i}{\sqrt{2m\omega_0}}, \ \ {\dot z}(0) = \sqrt{\frac{\omega_0}{2m}},
\label{zic}
\ee
and are equivalent to
\be
\xi(0)=0 , \ {\dot \xi}(0) =\sqrt{\frac{\omega_0}{2m}}\ ; \ 
\chi(0)= \frac{-1}{\sqrt{2m\omega_0}}, \ {\dot \chi}(0)= 0
\label{xichiic}
\ee
where $\omega_0 =\omega(0)$. (The expressions differ
from those in~\cite{Vachaspati:2018pps} in the factors of $\sqrt{2}$
and $m$ because of different conventions.)
The initial conditions and Eq.~(\ref{xichieq}) imply that the Wronskian is conserved,
\be
W \equiv \xi {\dot \chi} - \chi {\dot \xi} = \frac{1}{2m}
\ee
which can also be written as
\be
z^* \, p_z  - p_z^* \, z = i.
\ee
where $p_z \equiv m {\dot z}$.

The initial conditions have the simple interpretation that the 
two-dimensional cSHO initially has the same energy as the qSHO
in its ground state. This is easy to see because $\beta(0)=0$ 
in Eq.~(\ref{Eq}) gives $E_q(0)=\omega_0/2$. What is more novel
is that the initial conditions are such that the two-dimensional cSHO also has
angular momentum $L=mW=1/2$. Furthermore, the angular momentum,
equivalently the Wronskian, is conserved during the evolution.
If we think of the qSHO as the mode coefficient of a free scalar
quantum field, the initial conditions imply that each mode of the 
corresponding classical complex scalar field must carry a conserved 
non-zero global charge. 

The quantum dynamical problem has thus transformed into a classical 
evolution problem for {\it any} time-dependent frequency $\omega(t)$. To
emphasize this point we write the full time-dependent annihilation operator
in terms of classical solutions,
\ba
a(t) =  \frac{(p_z^* - i m \omega z^* )}{\sqrt{2m \omega}} a_0 +
           \frac{(p_z -  i m \omega z)}{\sqrt{2m \omega}} a_0^\dag
\ea
where $z$ denotes a classical solution with the initial conditions given above.
Thus we have a mapping between the quantum solution and the classical
solution.

Finally we re-express the quantum energy in Eq.~(\ref{Eq}) in terms of the 
$\xi$ and $\chi$ variables,
\ba
E_q &=& \frac{|p_z|^2}{2m} + \frac{m\omega^2}{2} |z|^2 \nonumber \\
&=&
 \left ( \frac{m}{2}{\dot \xi}^2 + \frac{m\omega^2}{2} \xi^2 \right )
+ \left ( \frac{m}{2} {\dot \chi}^2 + \frac{m\omega^2}{2} \chi^2 \right ) \nonumber \\
& \equiv& E_\xi + E_\chi .
\label{EqEclass}
\ea

To summarize: to find the energy in quantum excitations, we simply have to solve the 
classical problem in Eq.~(\ref{zeq}) with the initial conditions in (\ref{zic}) 
and then calculate $E_q$ using (\ref{EqEclass}).
This is the CQC, earlier derived in the functional
Schr\"odinger picture~\cite{Vachaspati:2018pps}, but derived here in the Heisenberg 
picture via Bogoliubov transformations.

We should insist on the fact that Eq. \eqref{zeq} along with the initial conditions \eqref{zic} is 
simply a rewriting of Eqs. \eqref{alphaeq} and \eqref{betaeq} (with the assiociated initial conditions 
$\alpha(0)=1$ and $\beta(0)=0$). Hence there is no leeway to map the quantum problem to a different 
classical problem, say that of only one cSHO, or two cSHOs with different initial conditions. In other words 
the mapping requires the existence of a conserved quantity (the angular momentum of a two-dimensional 
cSHO) whose value is set by the constraint \eqref{conserved}.

\section{Backreaction}
\label{backreaction}

In the quantum problem, the time-dependent frequency produces quantum
excitations and must backreact on the source responsible for the time
dependence. In many situations, especially in gravitational settings, the quantum
backreaction is difficult to calculate. However, the backreaction in the corresponding
classical problem is in principle straightforward to evaluate because the classical
equations of motion are known. If the classical equations are difficult to solve
analytically, we can always, in principle, solve them numerically.
We will now illustrate such a backreaction calculation
for a toy problem that can be solved completely. This will tell us if the 
solution using the CQC is a good approximation to the full quantum
solution.

Our toy model consists of two quantum degrees of freedom, $x$ and $z$,
where $x$ represents a particle rolling down a linear potential and $z$
represents a simple harmonic oscillator that couples to the rolling particle.
(The model has similarities to field theories used in inflationary
cosmology and to the ``bottomless'' potentials considered in
Ref.~\cite{Vachaspati:2002cr}.) The Hamiltonian for the system is
\be
H = \frac{p_x^2}{2M} -  M a x + \frac{p_z^2}{2m} + \frac{1}{2} m \omega_0^2 z^2
+\frac{\lambda}{2} x^2 z^2
\ee
which we shall rescale and write with re-defined $a$, $\omega_0$ and $\lambda$ 
as
\be
H = \frac{p_x^2}{2} -  a x + \frac{p_z^2}{2} + \frac{1}{2} \omega_0^2 z^2
+ \frac{\lambda}{2} x^2 z^2.
\label{ham}
\ee
Here $a$ corresponds to the constant
classical acceleration while rolling, $\omega_0$ is the simple harmonic
oscillator frequency in the absence of any coupling to the rolling particle, and
$\lambda$ is the coupling.


We are mainly interested in the dynamics of the rolling particle and how
the presence of the simple harmonic oscillator backreacts on the dynamics.
So we will first solve the classical rolling problem, then find the simple
harmonic oscillator solution in the ``fixed background'' approximation.
Next we will solve for the full dynamics using the CQC
described above. Finally we will solve the full quantum
problem and compare with the result obtained using the CQC.

\subsection{Classical solution}
\label{classicalsoln}

The classical equations of motion are
\be
{\ddot x} = a - \lambda x z^2, \ \ {\ddot z} = - (\omega_0^2+ \lambda x^2) z
\ee
If the initial conditions (at $t=0$) are
\be
x(0) = 0, \ \ {\dot x}(0)=0, \ \ z(0)=0, \ \ {\dot z}(0)=0
\ee
then the solution is
\be
x(t)=  \frac{1}{2}a t^2 , \ \ z(t)=0.
\ee
That is, the rolling particle continues to roll with constant acceleration
while the simple harmonic oscillator degree of freedom is not excited.

\subsection{Fixed background analysis}
\label{fixedbackground}

\begin{figure}
      \includegraphics[width=0.45\textwidth,angle=0]{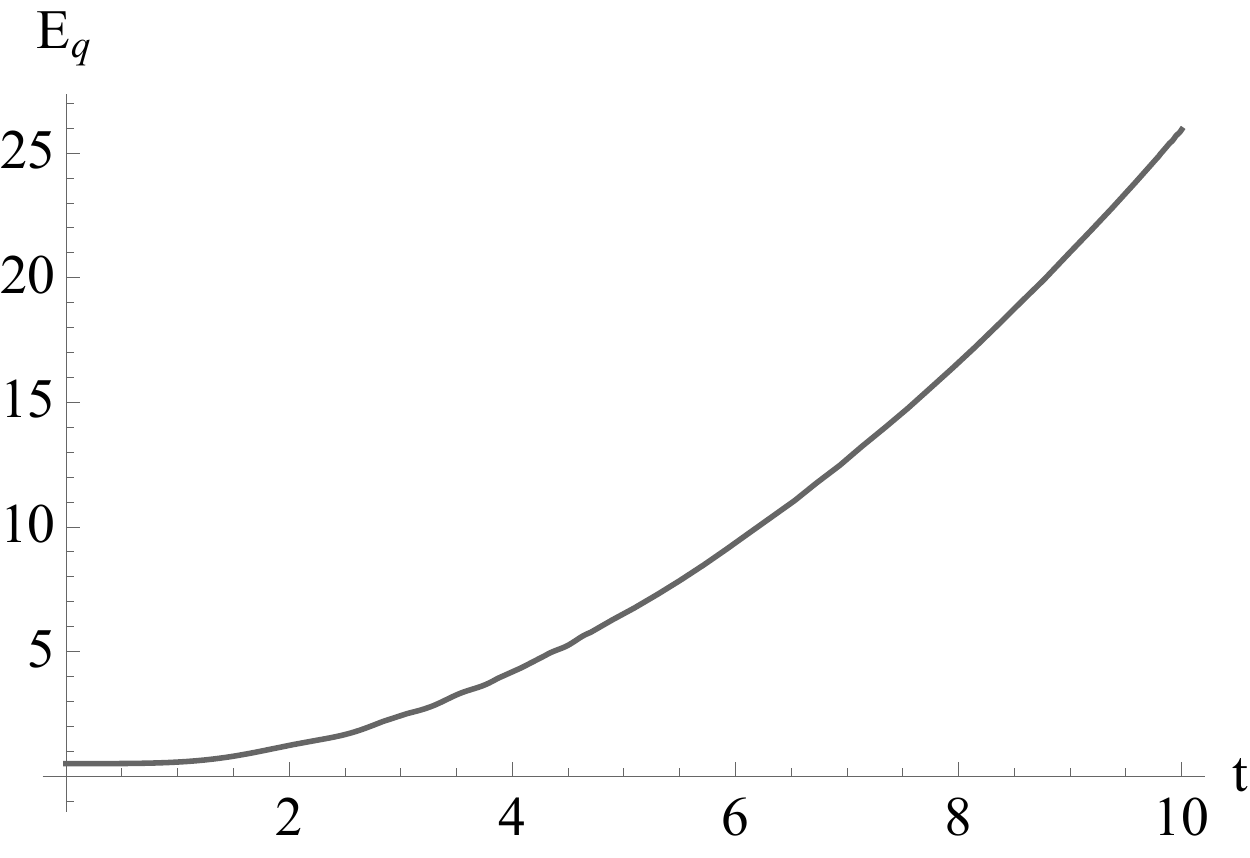}
  \caption{$E_q$ vs. $t$ in ``fixed background approximation'' for $\omega_0=1$,
  $a=1$. The energy of the background is conserved but the total energy is
  not conserved.
}
\label{EqVstFB}
\end{figure}

In the fixed background of the rolling particle, the CQC is exact and the Hamiltonian 
for the simple harmonic oscillator is
\be
H_z =  \frac{p_z^2}{2} + \frac{1}{2} \omega^2(t) z^2
\ee
where $z=\xi+i\chi$ and
\be
\omega^2(t) \equiv \omega_0^2+ \frac{\lambda}{4} a^2 t^4.
\ee
Then the energy of the simple harmonic oscillator can be found from
Eq.~(\ref{EqEclass}) where we need to solve the classical equations
of motion in Eq.~(\ref{xichieq}) with the initial conditions in Eq.~(\ref{xichiic}).
With $\omega_0=0$, Eq.~(\ref{xichieq}) can be solved in terms of Bessel
functions but for $\omega_0 \ne 0$ we have to resort to a numerical computation.
The result for $E_q(t)$ with $\omega_0=1$, $a=1$, $\lambda=1$ is shown in Fig.~\ref{EqVstFB}.
Note that total energy is not conserved in the fixed background analysis:
initially the energy is $\omega_0/2 = 0.5$ while at $t=20$ we see that it has
grown to $\sim 25$.

\subsection{Backreaction with CQC}
\label{cqbackreact}

To obtain dynamics with backreaction with the CQC,
we need to solve the classical equations
\ba
&& {\ddot x} = a - \lambda x (\xi^2 + \chi^2), \\
&&
{\ddot \xi} = - (\omega_0^2+ \lambda x^2) \xi, \\ 
&&
{\ddot \chi} = - (\omega_0^2+ \lambda x^2) \chi,
\ea
with initial conditions
\ba
&& x(0) = 0, \ \ {\dot x}(0)=0, \\
&& \xi(0)=0 , \ \ {\dot \xi}(0) =\sqrt{\frac{\omega_0}{2}}, \\
&& \chi(0)= \frac{-1}{\sqrt{2\omega_0}}, \ \ {\dot \chi}(0)= 0.
\ea
This system of equations is solved numerically. 

\begin{figure}
      \includegraphics[width=0.45\textwidth,angle=0]{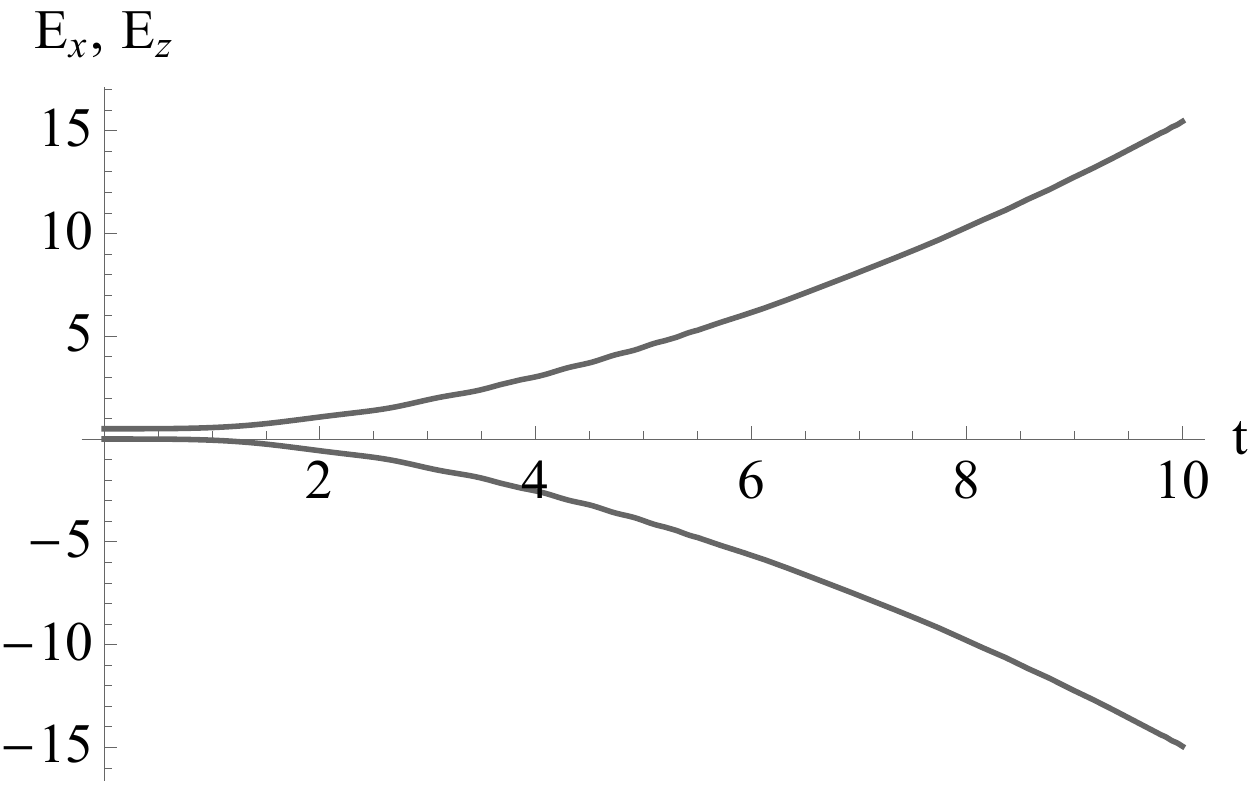}
  \caption{The energy in the simple harmonic oscillator versus time (upper curve)
  as calculated with the CQC. The lower curve
  shows the energy in the rolling particle. The interaction term
  $x^2z^2/2$ is included in the energy of the simple harmonic oscillator (upper curve). 
  The total energy is conserved.
}
\label{exezincquantum}
\end{figure}

In Fig.~\ref{exezincquantum} we show how the energy in the simple harmonic oscillator grows 
with time and that in the rolling particle decreases with time. The total energy is
conserved. We will show the solution for $x(t)$ using the CQC below, after we have
discussed the solution of the full quantum problem.

\subsection{Full quantum treatment}
\label{fulltreatment}

To solve for the full quantum dynamics, we have to solve the time-dependent
Schr\"odinger equation
\be
H \psi(x,z,t) = i \frac{\partial\psi}{\partial t}
\ee
with $H$ given in Eq.~(\ref{ham}).
The initial wavefunction is taken to consist of Gaussian wavepackets in both the
$x$ and $z$ variables,
\ba
\psi(t=0,x,z) &=&  \left ( \frac{1}{\pi \sigma_x^2} \right )^{1/4} e^{- x^2/(2 \sigma_x^2)}
\nonumber \\
&& \hskip 0.5 cm
 \times \left ( \frac{\omega_0}{\pi} \right )^{1/4} e^{-\omega_0 z^2/2}
 \label{icpacket}
\ea
The parameter $\sigma_x$ is a free parameter in the full quantum problem
and we shall study the dynamics for several values of $\sigma_x$.

With the initial condition in Eq.~(\ref{icpacket}), we have $\langle x \rangle =0$ 
at $t=0$ for all $\sigma_x$. Ehrenfest's theorem in the absence of backreaction 
($\lambda=0$) gives the classical result for the evolution of the expectation value 
of $x$,
\be
\langle x \rangle_{\lambda=0} = \frac{1}{2} a t^2.
\ee
We are interested in determining the effect of backreaction on this evolution.

Standard algorithms to solve the Schr\"odinger equation numerically, such as the explicit 
Crank-Nicholson method with two iterations, were found to be unstable. After some
experimentation we found that Visscher's algorithm described in the Appendix is
stable, provided we use a very small time step for the evolution. This limited the
duration for which we could evolve the system, though it is sufficiently long to test the 
CQC.
%
The numerical solution yields the wavefunction from which we then calculate the expectation 
value of the position of the rolling particle, $\langle x \rangle$. (Symmetry under $z\to -z$ 
gives $\langle z \rangle=0$ at all times.)




\begin{figure}
      \includegraphics[width=0.45\textwidth,angle=0]{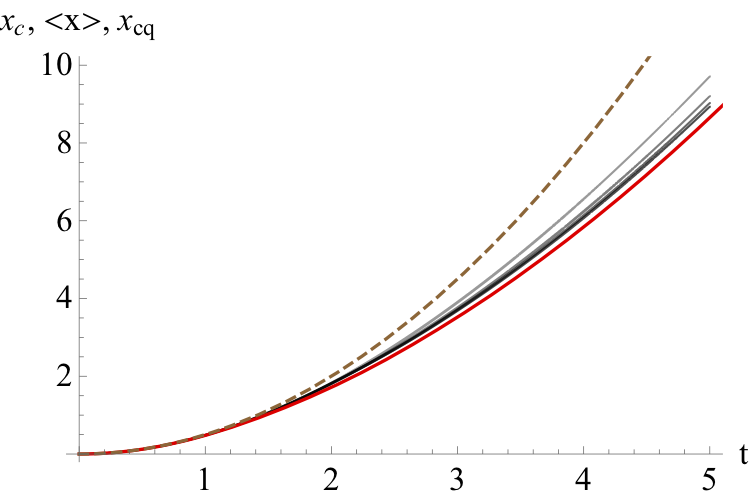}
  \caption{Rolling as calculated in the different analyses for $a=1$, $\omega_0=1$
  and $\lambda=1$. The dashed curve shows the classical solution, $x_c(t)=at^2/2$,
  and ignores backreaction. The gray curves show the rolling in the full quantum
  treatment with $\sigma_x=0.5, 1, 1.5, 2.0$, with the curves getting lower with
  increasing $\sigma_x$. The lowest (red) curve shows the rolling found using the CQC.
  }
\label{wxdependence}
\end{figure}


\begin{figure}
      \includegraphics[width=0.45\textwidth,angle=0]{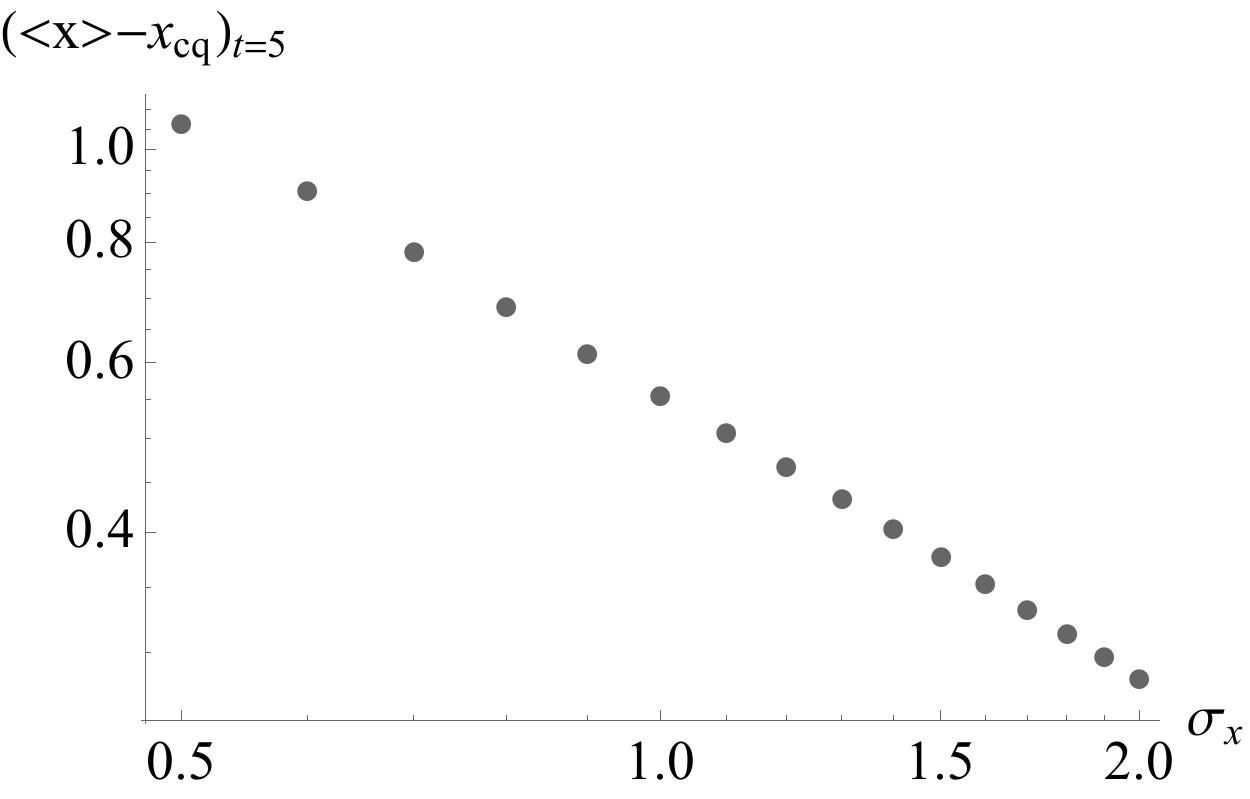}
  \caption{Log-log plot of $\langle x\rangle - x_{cq}$ at $t=5$ showing that the CQC
  becomes more exact for larger $\sigma_x$. }
\label{regression}
\end{figure}

In Fig.~\ref{wxdependence} we show the dynamics of the rolling particle in all the
different treatments: first the evolution ignoring backreaction, then
the full quantum calculation for several values of $\sigma_x$ where backreaction 
is automatically included,  and finally the evolution with backreaction evaluated 
using the CQC. 
It appears that the fully quantum treatment and the CQC agree to better and better accuracy as $\sigma_x$ grows larger. 
To quantify this phenomenon, we also plot the difference of quantum and CQC evolutions at a
fixed time ($t=5$) for different values of $\sigma_x$ (Fig. \ref{regression}). The fit to the line gives
\be
\langle x \rangle_{t=5} \approx x_{cq}(t=5) + \frac{2}{\sigma_x}.
\ee
Therefore the full quantum result goes to the CQC
result in the limit of large $\sigma_x$.

Since the wavepacket spreads during time evolution, we would also expect the agreement between CQC and 
fully quantum treatment to become exact at late times. To understand why this would be a reasonable expectation, 
we consider the time-dependent wavepacket solution for a {\it free} particle,
\be
\psi(t,x) = A(t,x) \exp \left ( - \frac{x^2}{2\sigma_x^2} \frac{1}{1+ \frac{t^2}{m^2\sigma_x^4}}
                    \right )\,,
\ee
where $A(t,x)$ is a complex number whose modulus does not depend on $x$. Note that the parameter $\sigma_x$ 
quantifies the initial width of the wavepacket and is proportional to the standard deviation on the initial position $x$ 
of the free particle. At late times the width of the wavepacket grows as
\be
\sigma(t) \sim \frac{t}{\sigma_x m}.
\ee
So the rate of wavepacket spreading is $ (\sigma_x m)^{-1}$. This wavepacket
spreading is a completely quantum effect. For a rolling particle to behave classically,
the rate of spreading should be much less than the rate at which it rolls,
\be
\frac{1}{\sigma_x m} \ll a t,
\ee
where $a$ is the constant acceleration of the particle.
Thus a rolling particle behaves more classically at late times, and the time
at which it starts behaving classically occurs earlier if the initial width of the
initial wavepacket is larger. Based on this behavior of a free particle,
as the rolling becomes more classical, the CQC 
should become more exact, and at late times it should match the quantum evolution.

\begin{figure}
  \includegraphics[width=0.45\textwidth,angle=0]{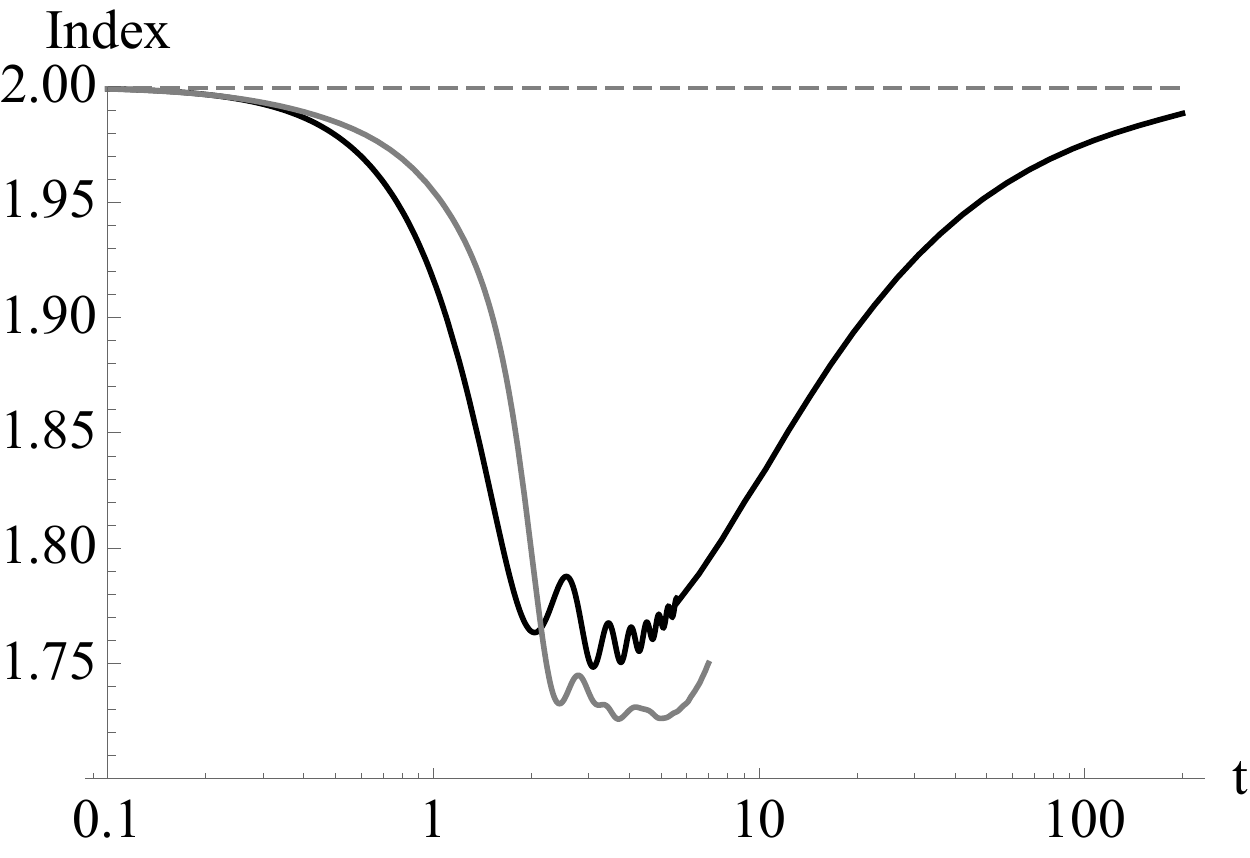}
  \caption{Log-linear plot of the scaling index 
  $n_s=d\ln(f)/d\ln(t)$ for $f=x_{cq}$ (in black) and $f=\langle x\rangle_{\sigma_x=2}$ (in gray). 
  At late times the scaling index of $x_{cq}$ approaches 2.}
\label{index}
\end{figure}

It is therefore instructive to study the late-time scaling behavior of the dynamics with backreaction of the 
rolling particle in the CQC (Fig.~\ref{index}). After an early transient phase where the particle's dynamics is 
damped and $x(t)\propto t^{1.75}$, the acceleration asymptotically approaches a constant ($x(t)\propto t^{2}$), 
albeit different from $a$. More precisely, the corresponding asymptotic solution is 
\ba
x(t)&=&\frac{1}{2}\left(a-\frac{\sqrt{\lambda}}{2}\right)t^2\,,\\
\xi(t)^2+\chi(t)^2&=&\frac{1}{\sqrt{\lambda}\left(a-\frac{\sqrt{\lambda}}{2}\right)t^2}\,.
\ea
Notice the peculiar fact that the latter equality requires the coupling to be bounded in order for this asymptotic solution 
to exist. (For $\lambda > 4$ we find bound state solutions~\cite{Vachaspati:2002cr}.)
This renormalization of the acceleration suggests that the vacuum effectively provides a
``quantum friction'' force of magnitude $-\sqrt{\lambda}/2$. In principle, this scaling behavior should also be recovered 
at late times in the full quantum 
treatment and this can be tested numerically in our toy-model. It turns out however that integrating the Schr\"odinger 
equation for longer periods of time will require a lot of computational power, or a more efficient and stable numerical
algorithm. For this reason we have only 
plotted the scaling behavior of the fully quantum $\langle x\rangle$ (for $\sigma_x=2$) up to $t=7$
in Fig.~\ref{index} and see good agreement with the CQC.
We leave a more thorough numerical analysis of the Schr\"odinger equation for future work.

\section{Conclusions}
\label{conclusions}

We have derived the CQC
in the Heisenberg picture. This shows that the dynamics of a quantum simple
harmonic oscillator with a time-dependent frequency is given by the dynamics 
of {\it two} classical simple harmonic oscillators with the same time-dependent
frequency and prescribed initial conditions. Equivalently, the quantum dynamics
can be recovered by complexifying the phase space variables of the classical 
simple harmonic oscillator. Since the modes of a free quantum 
field in a background can be treated as an infinite set of simple harmonic oscillators
with time-dependent frequencies, the CQC can be 
extended to field theory. Then the dynamics of a {\it quantum real} scalar field is 
given by the dynamics of a {\it classical complex} scalar field, again with 
prescribed initial conditions.

The CQC provides a tool to study the backreaction
of quantum radiation on classical dynamics. We have investigated the backreaction
in a toy model that involves a particle rolling down a linear potential and coupled
to other simple harmonic oscillator degrees of freedom. We solved this toy problem
using the CQC and compared it to the full quantum
solution. The dynamics in the two approaches agree remarkably well, especially
as the initial quantum state in the full treatment is taken to be more classical
({\it i.e.} larger $\sigma_x$).
Furthermore, the analysis using the CQC is trivial
to implement numerically, whereas the full quantum treatment is a non-trivial
numerical task.

\begin{figure}
      \includegraphics[width=0.45\textwidth,angle=0]{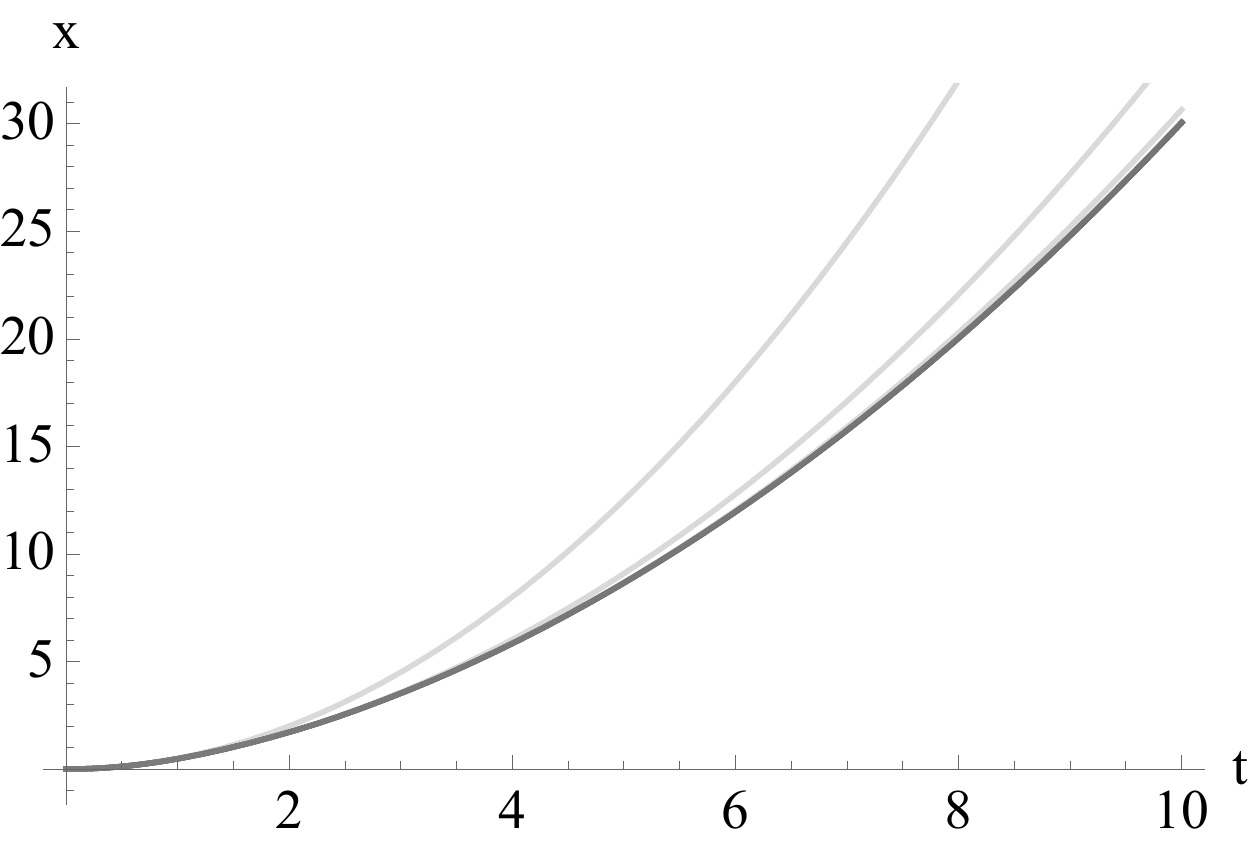}
  \caption{Background $x_N$ vs. $t$ for increasing number $N$ of iterations of the semiclassical procedure 
  (gray curves) approaching the CQC result (black curve) from above.}
\label{semiclassicalx}
\end{figure}

\begin{figure}
      \includegraphics[width=0.45\textwidth,angle=0]{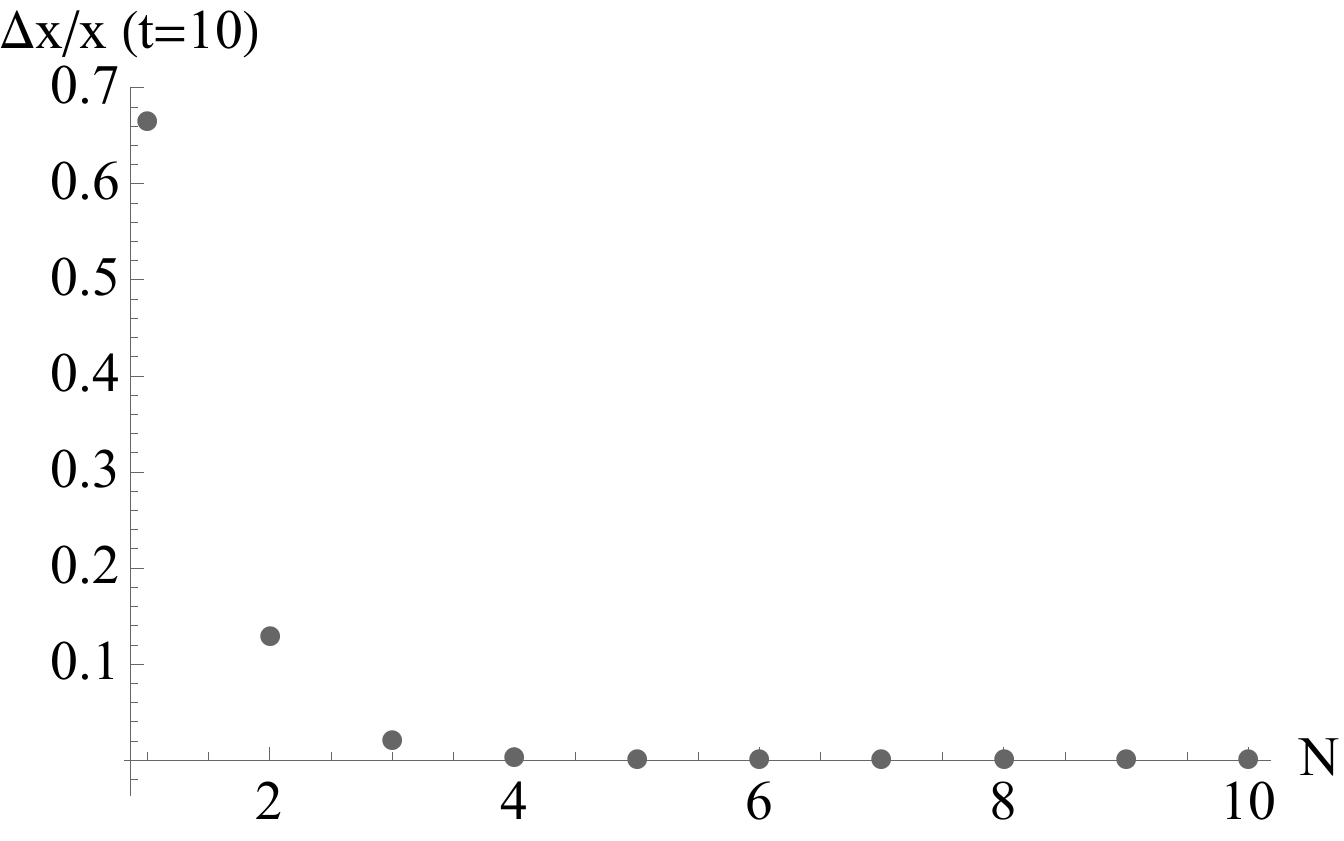}
  \caption{Fractional error $(x_N-x_{cq})/x_{cq}$ at time $t=10$ vs. number of iterations $N$.}
\label{semiclassicalDeltax}
\end{figure}

We can also compare the CQC with the iterative semiclassical approach to 
calculate the radiation and backreaction. There the radiation is calculated 
in a fixed background, then the background equations are solved using the
radiation solution. This modifies the background and the procedure can be
repeated in the modified background. In practice only a few iterations are
performed and it is assumed that the procedure will converge. To assess 
the effectiveness of this procedure, we plot the semiclassical background $x_N(t)$ 
after $N$ iterations of this procedure (Fig.~\ref{semiclassicalx}) as well as the fractional error 
when compared to the CQC result (Fig.~\ref{semiclassicalDeltax}). (Analogous plots can be 
obtained for the radiated energy.) We thus see that the iterative procedure converges quickly 
for  $t\leq10$, the fractional error being at the level of the working numerical precision after only 3 iterations. 
However as the time increases, more iterations will be needed to provide a good approximation to the 
full quantum problem (note that the relative error is $\sim 100\%$ after only one iteration). The CQC 
is therefore superior since it bypasses this iterative semi-classical procedure and gives the exact 
result for all times.

We expect the CQC to have wide applicability since quantum excitations on
classical backgrounds occur in many physical systems. The approach could
prove invaluable in the gravitational context where one considers quantum
fields in curved spacetime. Then the common approach is to work with
``semiclassical gravity''~\cite{Birrell:1982ix}, {\it i.e.} with the Einstein equation
modified to
\be
G_{\mu\nu} = 8 \pi G \langle T^{\rm ren}_{\mu\nu} \rangle
\ee
where $G_{\mu\nu}$ is the Einstein tensor. The right-hand side 
is the expectation value of the renormalized energy-momentum tensor of the 
radiation fields in a suitable quantum state -- this is the energy-momentum
tensor after the vacuum energy has been subtracted out and other bare
couplings have been adjusted to reduce the equation to the above form.
In principle semiclassical gravity and extensions may provide an iterative 
scheme for calculating the backreaction of quantum fields on the spacetime.
The CQC approach however is to solve the {\it classical} equations
\be
G_{\mu\nu} = 8 \pi G T_{\mu\nu}'
\ee
where the prime on the right-hand side denotes that it is the classical
energy-momentum tensor for the complexified fields minus the vacuum
energy contribution. (Depending on the physical situation of interest, we 
could include a cosmological constant term.)
This modified Einstein equation would then
be solved together with the classical field equations
\be
\nabla_\nu {T^{\mu\nu}}' = 0
\ee
with suitable initial conditions as discussed in this paper. The solution would
provide the complete time dependence of the fields as well as the spacetime. 
A successful analysis in the case of gravitational collapse promises to shed
light on black hole formation and the information paradox as already indicated
in Ref.~\cite{Vachaspati:2018pps}. 

\acknowledgements
We are grateful to several colleagues at the PASCOS 2018 meeting, especially
Mark Hertzberg, Harsh Mathur, Paul Saffin and Andrew Tolley, for feedback.
We are also grateful to Jan Olle Aguilera for a careful reading of the manuscript
and to Sean Bryan for useful discussions.
TV's work is supported by the U.S. Department of Energy, 
Office of High Energy Physics, under Award No. DE-SC0013605 at Arizona State 
University and GZ is supported by John Templeton Foundation grant 60253.\\

\appendix

\section*{Appendix: Numerical Method}
\label{numerics}

Although our toy model Hamiltonian appears simple, standard numerical
algorithms led to severe numerical instabilities. We eventually found the
simple but effective algorithm due to Visscher in Ref.~\cite{doi:10.1063/1.168415}
which worked for our problem, even though we had to use very small time steps
in the evolution.
The idea is to write the Schr\"odinger equation in terms of the real and imaginary
parts, $\psi_R$ and $\psi_I$, of the wavefunction
\be
\partial_t \psi_R = H \psi_I, \ \ \partial_t \psi_I = - H \psi_R.
\ee
The novelty is that $\psi_R$ is taken to be at integer time steps while $\psi_I$ is 
taken to be at half-integer time steps. The equations are then discretized in the 
usual way by replacing spatial derivatives by central differences. The time derivative
is also central which is seen for example by
\be
\psi_R(t+1,x)-\psi_R(t,x) = dt \times H\psi_I (t+1/2,x)
\ee
As the right-hand side is evaluated half way between the times at which the
differences on the left-hand side are evaluated, this gives 
second order accuracy in $dt$ and stability if $dt$ is small 
enough~\cite{doi:10.1063/1.168415}.

The probability density at any integer time step $t$ is given by
\be
P(t,x) = (\psi_R (t,x))^2 + \psi_I (t-1/2,x) \psi_I (t+1/2,x).
\ee
A good numerical check of the code is that the total probability should be unity
at all times and the total energy should be conserved. Expectation values of $x$ 
are calculated using this expression for the probability density. 

An estimate of the numerical error is obtained by evolving the Schr\"odinger
equation forward and then backward in time. The final result should give 
$\langle x \rangle =0$ (the initial condition). Half the deviation gives an estimate 
of the numerical noise error and was negligible ($\sim 10^{-7}$) in our check.

\bibstyle{aps}
\bibliography{classicalBogoliubov}

\end{document}